\documentclass[aps, prd, twocolumn, lengthcheck, superscriptaddress, 
nofootinbib]{revtex4-1}

\usepackage{epsfig}
\usepackage[usenames]{color}
\usepackage{graphicx}
\usepackage{amsmath}
\usepackage{epstopdf}

\newcommand\sect[1]{\emph{#1.}---}
\def\bi{\bibitem}

\def\be{\begin{eqnarray}}\def\ee{\end{eqnarray}}
\def\lsim{\mathrel{\rlap{\lower3pt\hbox{\hskip1pt$\sim$}}
     \raise1pt\hbox{$<$}}} 
\def\gsim{\mathrel{\rlap{\lower3pt\hbox{\hskip1pt$\sim$}}
     \raise1pt\hbox{$>$}}} 

\allowdisplaybreaks

\begin{document}

\title{The Quenched  ${g_A}$ in Nuclei and
  Emergent Scale Symmetry  in Baryonic Matter}

\author{Yong-Liang Ma}
\email{yongliangma@jlu.edu.cn}
\affiliation{Center for Theoretical Physics and College of Physics, Jilin University, Changchun, 130012, China}

\author{Mannque Rho}
\email{mannque.rho@ipht.fr}
\affiliation{Universit\'e Paris-Saclay, CNRS, CEA, Institut de Physique Th\'eorique, 91191, Gif-sur-Yvette, France }

\date{\today}

\begin{abstract}
 The recent RIKEN experiment on the quenched $g_A$ in the superallowed Gamow-Teller transition from $^{100}$Sn indicates a role of scale anomaly encoded in the anomalous dimension $\beta^\prime$ of the gluonic stress tensor ${\rm Tr \ G}_{\mu\nu}^2$.  This observation provides a support to the notion of hidden scale symmetry emerging by strong nuclear correlations with an IR fixed point realized --- in the chiral limit --- in the Nambu-Goldstone mode. We suggest there is an analogy in the way scale symmetry manifests in nuclear medium to the continuity from the unitarity limit at low density (in light nuclei) to the dilaton limit at high density (in compact stars). In between the limits, say,  at normal nuclear matter density, the symmetry is not visible, hence hidden.
\end{abstract}

\maketitle

\sect{ Introduction}
There is a long-standing ``mystery" lasting more than four decades~\cite{wilkinson} as to why the Gamow-Teller (GT) transition in simple shell model in nuclei requires a quenching factor $q\sim (0.75 - 0.80)$ multiplying the axial coupling constant $g_A^{\rm free}=1.276$ which would make $g_A^{\rm eff} \to 1.0$ (see, e.g., Ref.~\cite{review-gA} for extensive up-to-date reviews). The standard nuclear $\beta$ decay process typically involves a super-allowed transition with zero momentum transfer, so it was natural to associate $g_A^{\rm eff}\simeq  1$ with something more {\it fundamental} than standard nuclear many-body interactions, such as {\it basic}  renormalization due to the vacuum change induced by the nuclear medium. That the effective $g_A^{\rm eff}$ involving a nearly conserved axial current is near unity reminded one --- falsely as is now understood --- of the CVC hypothesis where $g_V=1$. The question then arose as to whether this constant $g_A^{\rm eff}$ near 1 could signal  certain thus-far unrecognized intrinsic properties of the underlying theory currently accepted,  QCD,  or just a  coincidental outcome arising entirely from mundane strong nuclear correlations or from a combination of both fundamental and mundane. This is an important question, not just for nuclear physics but also for going beyond the Standard Model, given that the {GT} matrix element figures importantly in neutrinoless double beta ($0\nu\beta\beta$) decays involving non-negligible momentum transfers, hence not supper-allowed.

We discuss in this note how the quenching of $g_A$ in nuclei and dense matter reveals the way scale symmetry, hidden in the vacuum in QCD,  manifests through strong nuclear correlations and make a conjecture on its implication on the {IR} fixed-point structure of QCD.

The approach most appropriate to address this issue is effective field theories (EFT) ``modeling"  QCD. The scheme currently adopted by most of the theorists in nuclear physics resorts to a cutoff $ \sim (400- 500)$ MeV, giving what's now established  as ``standard chiral effective field theory" (S$\chi$EFT for short) where only the nucleons and pions figure as the relevant degrees of freedom.  As described in great detail in Ref.~\cite{MR-PCM}, we find it far more powerful and predictive to resort to a higher cutoff,  $\gsim 700$ MeV, in particular for going to high densities relevant to massive compact stars. The relevant degrees of freedom with such a cutoff are the lowest-lying vector mesons $V_\mu =(\rho$,   $\omega$)  with mass $\gsim 700$ MeV possessing ``hidden local symmetry (HLS)" and a scalar meson with mass $\sim 600$ MeV denoted $\chi$ as a dilaton { --- pseudo-Nambu-Goldstone boson --- } of broken scale symmetry.
 The model consisting of $V_\mu$ and $\chi$ together with the nucleons constructed in consistency with both scale and chiral symmetries will be referred to as  ``generalized scale-chiral  EFT" (G$\sigma$EFT). It has been established that this G$\sigma$EFT in what is referred to as a ``leading order scale symmetry (LOSS)" approximation is surprisingly  successful, with very few parameters,  for describing not only nuclear matter at the equilibrium density {$n_0\simeq 0.16$ fm$^{-3}$} but also the compact-star matter at $n\sim (5-7)n_0$~\cite{MR-PCM}.

In the same LOSS approximation, it has been shown that the quenching factor in $g_A^{\rm eff}\approx 1$ is given {\it predominantly}, if not entirely,  by standard nuclear correlations, with little corrections from intrinsic QCD effects\footnote{These include multibody meson-exchange currents that are counted as higher-order corrections in S$\chi$EFT.}.  Any significant  deviation from $g_A^{\rm eff}\approx 1$ would then have to be considered as a signal for scale-symmetry (explicit) breaking, a quantum anomaly, in terms of the anomalous dimension $\beta^\prime$ of the gluonic stress tensor ${\rm Tr}\ G_{\mu\nu}^2$. This issue of possible deviation from $g_A^{\rm eff}\approx 1$ becomes particularly relevant and poignant in nuclear physics --- and possibly in $0\nu\beta\beta$  processes --- due to upcoming precisely measured {GT} transitions in doubly magic nuclei where the  ``extreme single particle shell model (ESPM)" is applicable.
This will be the principal issue treated in this paper.

\sect{What does Nature say?}
We first summarize what we consider to be a relevant global indication in the observed GT transitions.  For this we arbitrarily consider two nuclear mass regions: $A \lsim 60$  (light)  and $A> 60$ (heavy). Among the many available, we resort to Ref.~\cite{review-gA}.

In nuclei up to $A\sim 60$, $g_A^{\rm eff}$ in shell model comes out to be
\be
g_A^{\rm eff}=q_{\rm light}  g_A^{\rm free}\approx 0.98 -1.18\label{gA=1}
\ee
with $g_A^{\rm free}=1.276$ given by the neutron decay in matter-free vacuum. {For reasons clarified later, we simply give only the relevant ranges, eschewing error bars,  in theoretical estimates.} In the range $q=0.76 - 0.93$ implied by (\ref{gA=1}), let us pick what gives $g_A^{\rm eff} \approx 1$
\be
q_{\rm light}\approx  0.78.\label{q-gA=1}
\ee

As the mass number goes up above $A\sim 60$, it is seen that the quenching factor tends to decrease to $q\sim 0.5$ for,  e.g., $A\sim 100$.

It should be stressed here that what one obtains in the shell-model calculations depends on details of the model space, correlations included  etc. In stead of the specific values in light nuclei listed in Ref.~\cite{review-gA}, what matters is the rough value  for $g_A^{\rm eff}$  near unity and the decreasing tendency with the increasing mass number. The calculation we will rely on,  described below,  is made in the  G$\sigma$EFT, resorting to a Fermi-liquid fixed point {(FLFP)} theory. Now the question is
which shell model calculation maps to the G$\sigma$EFT result? Our proposal is the super-allowed GT decay of the doubly magic nucleus $^{100}$Sn.\footnote{This process has been exploited in Ref.~\cite{firstprinciple} to argue for a ``first-principle resolution" of the $g_A$ quenching problem.  We contest this point below.}
What is crucial here is that it provides the ESPM~\cite{Sn100} to be identified with the result of G$\sigma$EFT.

Briefly stated, the doubly magic $^{100}$Sn has 50 neutrons and 50 protons in completely occupied states. In an ESPM description, the GT transition  involves  the decay of a proton in a completely filled shell ($g_{9/2}$) to a neutron in an empty shell ($g_{7/2}$), thereby giving a precisely defined quenching factor $q^{\rm ESPM}$.
Limited to the configuration space of the nucleons only, the $q_{\rm ESPM}$ completely captures nuclear correlations up to but below the $\Delta$-N mass difference.
In the FLFP description exploited here, the transition corresponds precisely to the GT transition of a quasi-proton  to a quasi-neutron  on top of the Fermi sea with the quenching factor denoted as $q_{\rm FL}$, implying that the renormalization-group $\beta$ function for the quenching factor in the {FLFP} theory is nearly zero.  Therefore we arrive at  the relation
\be
q_{\rm FL}\simeq q^{\rm ESPM}.
\ee

\sect{Superallowed GT decay in $^{100}$Sn}
There is at present a conflicting information on the superallowed transition in $^{100}$Sn. This generates an interesting future development for the fundamental issue of scale symmetry in nuclear medium.

In the old GSI measurement~\cite{Sn100}, the GT strength was
\be
{\cal B}_{\rm GT}^{\rm GSI}=9.1^{+2.6}_{-3.0}.
\ee
In the ESPM~\cite{espm}, the GT strength for the $^{100}$Sn decay to $^{100}$In comes out to be~\cite{espm-sn100}
\be
{\cal B}_{\rm GT}^{\rm ESPM}= 17.78.\label{espsB}
\ee
This gives the quenching factor
\be
q^{\rm ESPM}_{\rm GSI }\approx 0.6 - 0.8.
\ee
Given the large error bars, one can only say that this is not  inconsistent with Eq.~(\ref{q-gA=1}).

However there is a more stringent recent experiment from RIKEN that gives a lower GT strength with much less error bars~\cite{RIKEN}
\be
{\cal B}_{\rm GT}^{\rm RIKEN}=4.4^{+0.9}_{-0.7}
\ee
giving the quenching factor in the range
\be
q^{\rm ESPM}_{\rm RIKEN}= 0.46 - 0. 55.\label{RIKEN-EXP}
\ee
This is clearly at odds with $q_{\rm light}$ (\ref{q-gA=1}).

We now suggest  this difference offers a glimpse into how scale symmetry manifests in dense baryonic matter.

\sect{Prediction in G$\sigma$EFT }
This calculation was first done in~\cite{friman-rho} with a model that incorporates chiral-scale symmetry~\cite{br91} in the Landau-Migdal Fermi-liquid approach to nuclei~\cite{migdal}.  It has been shown recently~\cite{LMRgA} that the result obtained in~\cite{friman-rho} corresponds precisely to what one obtains in the LOSS approximation in G$\sigma$EFT~\cite{MR-PCM}, was in agreement with (\ref{q-gA=1}).

If the RIKEN result (\ref{RIKEN-EXP}) is reconfirmed in the future experiments, this would raise a question on the validity of not only the LOSS approximation in the analysis of compact-star properties made in \cite{MR-PCM} but also all other EFT approaches. This is because what is involved is the role of scale symmetry breaking associated with a quantum anomaly in the EoS for baryonic matter at all densities.  At present there is no nonperturbative calculation of the anomalous dimension $\beta^\prime$ which figures crucially in the scale anomaly of QCD, so this may give the first glimpse of how scale symmetry emerges in nuclear medium and in strong interactions in general.

To address this problem in the framework of G$\sigma$EFT, we write the quenching factor  as
\be
q^{\rm ESPM}_{\rm G\sigma EFT}=q_{\rm  ssb} \times q_{\rm snc}.\label{q-geft}
\ee
Here the $q_{\rm ssb}$ represents the quenching factor inherited from QCD due to explicit scale symmetry breaking (SSB) in G$\sigma$EFT, and $q_{\rm snc}$ accounts for strong nuclear many-body correlations (SNC).

\sect{Scale Symmetry in the Nuclear Axial Current}
Scale symmetry or more generally conformal symmetry in strong interactions is a highly controversial issue dating from 1960s, and even now there is no consensus among the experts.
The scale symmetry we are concerned with here is an {\it emerging}, not intrinsic, symmetry at low energy and not directly tied to the basic structure of scale anomaly. In nuclear physics, there is no question that a scalar degree of freedom, which may be identified with the $f_0(500)$ in the particle booklet, is essential, i.e.,  attraction in nuclear potentials and covariant density functional approaches, the existence of soft scalar modes at high density etc. At not too high densities, say $\sim n_0$,  S$\chi$EFT without a scalar has the power to generate the necessary scalar property when treated in high orders in chiral expansion involving pion fields only.
But it is highly doubtful that it can correctly capture the properties of scalar excitations, e.g., soft modes,  at high density as in compact stars.

Currently, scale symmetry figures prominently in Higgs physics for going beyond the Standard Model involving a large number of flavors~\cite{MR-PCM} where a narrow-width scalar ``meson"  with mass lower than pseudo-scalars appears. The scale symmetry we are concerned with in nuclear physics, however, involves $N_f\sim 2-3$. There is no such narrow-width scalar in QCD in the vacuum, so in some circles, it has been argued that there is no IR fixed-point structure in nuclear physics.

Whether there is an IR fixed point in QCD awaits nonperturbative calculations which for the moment remain difficult to perform. The problem we are dealing with here involves emergent symmetries that may not necessarily be directly connected to the fundamental symmetries of QCD. The issue concerns both HLS and scale symmetry which we will focus on.


It has been observed that at high densities relevant to compact stars, say, $n\gsim 3n_0$, there arise a variety of (approximate) symmetries induced by nuclear correlations~\cite{MR-PCM}. Most notable is the topology change  where a solitonic baryon, i.e., skyrmion, fractionalizes into two half-skyrmions with the chiral condensate averaged to zero but with non-vanishing pion and dilaton decay constants, $f_\pi\approx f_\chi\neq 0$, thus remaining in the Nambu-Goldstone (NG) mode. There also emerges  parity doubling in the baryon spectrum.
And in the dilaton limit, the fundamental axial coupling $g_A$ tends to unity.  Thus the IR structure is quite similar to the scheme proposed by Crewther and Tunstall (CT)~\cite{CT} in QCD and  in technicolor~\cite{CT,crewther}. In our case we are away from the possible IR fixed point if it exists, but what is relevant is that soft theorems with both the pseudoscalars $\pi$ and scalar $\chi$ govern the dynamics. It should be stressed that we have here scale symmetry emergent in nuclear correlations, not necessarily reflecting on fundamental. A notable possibility here is that {\it given that nuclear correlations are governed by QCD, one could relate the IR structure we have to what CT proposed for QCD~\cite{CT}.}

 This scheme leads to certain predictions that are not shared by other theories, e.g., density functional approaches, S$\chi$EFT etc~\cite{MR-PCM}. A notable case is the precocious onset of  conformal sound velocity at non-asymptotic density in compact stars. Our suggestion is that the $g_A$ quenching phenomenon is also consistent with the emerging symmetries as in dense baryonic matter.

 Following  Ref.~\cite{LMR-CSEFT} where the G$\sigma$EFT was worked out, we can write the axial current {\it in medium} in the form
  \be
 q_{\rm ssb}  g_A\bar{\psi}\tau^\pm \gamma_\mu\gamma_5\psi\label{axialcurrent}
 \ee
 with
 \be
 q_{\rm ssb}= c_A +(1-c_A)\Phi^{\beta^\prime}
 \ee where  $\beta^\prime$ is the anomalous dimension of ${\rm Tr}\ G_{\mu\nu}^2$ and $0\leq c_A\leq 1$ is an arbitrary constant. In the vacuum, $\Phi=1$, so the $\beta^\prime$ dependence is absent. In Nature the scalar mass is nonzero, so $\beta^\prime$ cannot be zero.  Now both $c_A$ and $\beta^\prime$, known neither empirically nor theoretically,  can be density-dependent in medium. On the other hand, {in medium}, the quantity $\Phi$ is defined by
 \be
 \Phi(n)=\frac{f_\chi^\ast (n)}{f_\chi}\simeq \frac{f_\pi^\ast(n)}{f_\pi} < 1 \ {\rm for} \ n\neq 0,
\ee
where $f_\chi$ ($f_\chi^\ast$) and $f_\pi$ ($f_\pi^\ast$) are, respectively, the dilaton  and pion decay constants in the vacuum (in medium). $f_\pi^\ast$  is known up to nuclear matter density by experiment~\cite{yamazaki}.

It is notable that the property of scale symmetry breaking in the GT operator appears entirely in the factor $q_{\rm ssb}$.  It can be simply associated  with an intrinsic QCD effect distinct from mundane nuclear correlations. Note that the $\beta^\prime$ representing scale anomaly, an explicit breaking, can figure with density dependence only when $c_A < 1$.

\sect{Quenching factor in the LOSS approximation}
In Ref.~\cite{LMRgA}, the LOSS approximation exploited in Ref.~\cite{MR-PCM} for compact-star physics (namely in the EoS) was invoked,
\be
c_A=1.
\ee
Hence in LOSS,
\be
q^{\rm LOSS}_{\rm ssb}=1.
\ee
Since in this approximation the $\beta^\prime$ effect is entirely lodged in the dilaton potential (giving mass to the scalar dilaton),   the current is simply  $g_A\bar{\psi}\tau^\pm \gamma_\mu\gamma_5\psi$.

The $q_{\rm snc}$ was worked out in Ref.~\cite{LMRgA} using the FLFP formula~\cite{friman-rho}. It has been argued~\cite{MR19} that the many-body meson-exchange currents figuring at N$^n$LO for $n\geq 3$ could be dropped.  The reason for this is as follows: The leading multi-body correction to the single-particle GT operator appears only at N$^3$LO~\cite{parketal}, consisting of a large number of terms --- more than $11$ terms~\cite{krebs} --- some of which are with unknown parameters. Furthermore there are also ``recoil terms" due to the inevitable non-relativistic approximation which are mostly ignored in the field. These are of the comparable strength to those terms taken into account. There can be  considerable cancelations among the terms as is noted in light nuclei~\cite{wiringa}. Hence unless all are included there is no reason to believe that the sum of part of the terms can give a reliable estimate. If those terms of N$^3$LO partially included are important as in Ref.~\cite{firstprinciple}, then the terms of N$^n$LO for $n>3$ must be included to be consistent with the chiral expansion, {which appears most likely impossible in practice}. In fact the ``chiral filter mechanism"~\cite{MR19} states that many-body  corrections to the GT operator be suppressed.  In Ref.~\cite{friman-rho}, those terms together with contributions from higher baryon resonances not figuring in the relevant degrees of freedom considered  are dropped for consistency with the Landau Fermi-liquid structure. This strongly suggests that the ``first-principle resolution" to the quenched $g_A$ in $^{100}$Sn made in Ref.~\cite{firstprinciple} is totally unfounded.

Now taking  the large $N_c$ and large $\bar{N}$ limits where $N_c$ is the number of colors and $\bar{N}$ is $k_F/(\Lambda-k_F)$ --- where $\Lambda$ is the cutoff in the momentum space of the Fermi sphere,  it was shown that~\cite{friman-rho}
\be
q^{\rm Landau}_{\rm snc}=\Big(1-\frac 13 \Phi \tilde{F}_1^\pi\Big)^{-2}\label{fixedFL}
\ee
where $\tilde{F}_1^\pi$ is the pionic contribution, precisely calculable by soft-pion theorems, to the Landau parameter $F_1$.

Now a crucial question is how accurate (\ref{fixedFL}) is and what are the possible corrections to it? Nobody knows how to compute $1/N_c$ correctios to $g_A$.  Here we assume that the leading approximation is good to the extent that low-energy theorems involving $g_A$ such as the Goldberger-Treiman relation are accurate. The large $\bar{N}$, more appropriate for the problem, however is a different matter. $1/\bar{N}$ which  is of order $1/3$ at normal nuclear matter density,  is not small. However that the FLFP approximation works remarkably well  for such low-energy EW processes  as the anomalous orbital gyromagnetic ratio $\delta g_l$ and the enhanced axial charge transition expressed in $\epsilon_{MEC}$, both in Pb nuclei,  as shown in \cite{friman-rho}, suggests  that the same could hold for the GT transition.  One can understand  this as that the FLFP approach captures certain (very) high order effects of the standard chiral EFT that suppress high-order $1/N_c$ terms. This is indicated by that the standard chiral EFT at the orders so far studied badly fails to explain those two processes. This aspect of the problem could be checked by doing systematic high-order $1/\bar{N}$ calculations in the $V_{lowk}$-renormalization-group formalism developed at Stony Brook. It is that formalism that was  applied to compact-star matter in \cite{MR-PCM} where all the relevant references can be found. This is in progress for both normal nuclear matter and dense compact-star matter. And, the work, including a nontrivial expansion in scale symmetry,  is also in progress.

With the value $\Phi (n_0)\approx 0.8$ from \cite{yamazaki},  we get
\be
q^{\rm Landau}_{\rm snc}\simeq 0.79\label{FL}.
\ee
Thus in the LOSS approximation
\be
g_A^{\rm Landau}\simeq 1.0.\label{gA1}
\ee
It turns out that Eq. (\ref{fixedFL}) is very weakly dependent on density, thus when  calculated at nuclear matter density in Fermi-gas model,  it is good for both light and heavy nuclei.

\sect{Impact of the anomalous dimension $\mathbf{\beta}^\prime$}
The SNC result (\ref{FL}), identified as the effect of nuclear correlations obtained in the LOSS without explicit $\beta^\prime$ dependence, does not imply that $\beta^\prime$ is negligible. The $\beta^\prime$ is responsible for  the scalar mass $m_\chi$ which is important in nuclear interactions, e.g., scalar-exchange  potential, so it cannot be zero. Now the deviation from the LOSS approximation arises from the $c_i$ coefficients with $c_i < 1$.

That the LOSS seems to work well in the EoS for nuclear matter in the G$\sigma$EFT approach~\cite{MR-PCM} could be taken as an indication for  $c_i\approx 1$. If however dense matter is described in HLS Lagrangian put on crystal it was realized that~\cite{omega} unless the $c$ coefficient in the homogeneous WZ (hWZ) term is $c_{\rm hWZ} < 1$, there can be no chiral transition at high density. The energy of nuclear matter at high density is found to diverge unless the effective $\omega$ mass goes to infinity which is totally at odds with Nature. This is directly related to the close interplay between the $\omega$-nucleon coupling in many-nucleon system phrased in G$\sigma$EFT, which is intimately connected with the scalar attraction~\cite{N-omega-interplay}.

To have an idea how things go,
let's assume $c_A\approx 0.15$ and $\beta^\prime\approx 2.0$ --- the same values that resolve the hWZ problem in Ref.~\cite{omega}.
That would give
\be
q_{\rm ssb}=c_A +(1-c_A)\Phi^{\beta^\prime}\approx 0.63
\ee
which leads to
\be
q^{\rm ESPM}_{\rm G\sigma EFT} = q_{\rm ssb} \times q_{\rm snc}^{\rm Landau} \approx 0.63\times 0.79\approx  0.50.
\ee
This could explain  the RIKEN result (\ref{RIKEN-EXP}) if the RIKEN data turns out to be  correct.

Now the obvious question is this. In nuclear as well as compact-star matter, the LOSS approximation with $c$ coefficients set equal to 1 fares well with Nature~\cite{MR-PCM}. Will the $c$ coefficients much  less than 1 as seem required in the hWZ and $g_A$ problems upset those ``good" results?  { An old analysis using a different formalism for incorporating the anomalous dimension indicates that there is no difficulty for nuclear matter ~\cite{song2}. Whether this is also the case in the CT scheme and in particular at high density where the LOSS approximation is more justified is being investigated~\cite{wen}.}

\sect{Continuity in scale symmetry from  dilute to dense baryonic matter}
There is an indication in nuclear observables that scale symmetry or conformal symmetry is present in nuclear interactions both at very low energy and density and very high density but in between there is no indication for such symmetry. At very low energy and density, at the unitarity limit (in the framework of pionless EFT),  conformal symmetry emerges  in light nuclei and in the EoS of baryonic matter~\cite{unitarity}. At the {normal nuclear matter density}, on the contrary, such symmetry is evidently absent, but at high density approaching the dilaton-limit fixed point (DLFP), the symmetry reappears~\cite{MR-PCM}. The precocious convergence to conformal sound velocity $v_s^2=1/3$ signals the emergence of the symmetry at high density.

A continuity between the two limits seems  to appear also in the  way $q_{\rm ssb}$ manifests in the $g_A$ quenching phenomenon going from light nuclei to heavy nuclei.  This would require that the quenching factor $q_{\rm ssb}$ be near 1 making $g_A^{\rm eff}\to 1$ in low-mass nuclei but drop to $\sim 1/2$ as one goes to heavier nuclei. This would be consistent with the observation in Ref.~\cite{review-gA}. At high density, $n\gg n_{1/2}$ where $n_{1/2}$ is the density at which the half-skyrmion phase sets, the approach to the DLFP ``forces"  the ``fundamental" $g_A \to 1$~\cite{MR-PCM}. So at the two limits, scale symmetry appears to manifest with $g_A^{\rm eff}\to 1$. We have no idea how this changeover takes place. It could perhaps involve a totally unexplored transition such as what's discussed in \cite{kan} involving topology with HLS and hidden scale symmetry.

\sect{Conclusion and remarks}
The G$\sigma$EFT formalism we developed before for the EoS of baryonic matter including massive compact-star matter~\cite{MR-PCM} indicates that the very recent data on the GT strength from RIKEN experiment~\cite{RIKEN} could expose how the scale symmetry hidden in the matte-free vacuum emerges  in the heavy nuclei sector. What seems to play a key role here is a non-negligible anomalous dimension $\beta^\prime$ of the gluonic stress tensor $\rm Tr\ G_{\mu\nu}^2$.  This observation suggests an analogy to the continuity from the unitarity limit at low density (in light nuclei) to the dilaton limit at high density (in compact stars). At both ends, there is an insinuation, albeit indirect, for a scale (or conformal) invariance. This is consistent with the observations made at some high density, say, $\gsim 3n_0$, namely, the possible vanishing of the (averaged) chiral condensate with however a non-vanishing pion (and dilaton) decay constant, parity doubling in the baryon structure and  pseudo-conformal sound velocity of the compact star, all consistent, up to today, with what's available in nature. Intriguingly the quenching of $g_A$ in light and heavy nuclei and in the dilaton limit obtained in G$\sigma$EFT, although it is not feasible to make a rigorous argument at present,  suggests, along with compact-star physics, the presence of an IR structure with the non-vanishing $f_\pi$ and $f_\chi$ at the IR fixed point, hence in the NG mode~\cite{CT}.
However given that nuclear correlations --- that give $q_{\rm snc}$ leading to $g_A^{\rm eff}\approx 1$ in light nuclei --- are dictated by QCD, the IR structure indicated in the emergent scale symmetry we discussed in this paper could very well support the CT scenario for QCD proper. Clearly a lattice measurement for the IR structure for $N_f \sim 2-3$ stressed by CT would be highly desirable.

\sect{Acknowledgments}
The work of YLM was supported in part by the National Science Foundation of China (NSFC) under Grant No. 11875147 and 11475071.

{\bf Note added}: 
While the manuscript was being refereed,  there took place a new development highly relevant to our principal thesis that  hidden scale symmetry  and also hidden local symmetry should emerge in baryonic matter via strong nuclear correlations regardless of whether they are intrinsic symmetries of QCD. It concerns the  observation in \cite{nature-physics} of what are considered to be ``deconfined quarks"  in the core of massive neutron stars, possibly evidencing the conformal sound speed $v_s^2 = 1/3$ and the polytropic index approaching $\gamma\approx 1$ characteristic of  ``quark degrees of freedom." In our G$\sigma$EFT approach, those are precisely the fractionally charged ``quasiparticles" giving rise to the pseudo-conformal sound velocity $v_s^2\approx 1/3$ and $\gamma\approx 1$ in the core of massive compact stars that have been recently predicted  in \cite{arxiv2020} based on the G$\sigma$EFT~\cite{MR-PCM}. Our proposition of this paper is that the $g_A=1$ at the dilaton-limit fixed point density, supposedly $\gsim 25n_0$, could be {\it linked} by quark-hadron continuity to the Landau fixed-point quantity $g_A^{\rm Landau}=1$ at normal nuclear matter density predicted in this paper for superallowed GT transitions in finite nuclei.  A possible deviation from $g_A^{\rm Landau}=1$ discussed in this paper, if further observed  and confirmed by precision analyses -- in the ESPM of future precision experiments, would then signal the role of the scale anomaly $\beta^\prime > 0$, so far totally elusive in lattice QCD calculations for $N_f \lsim 3$. Nature is of course never perfect, so this will give a first indication of how scale symmetry breaks down in nuclear physics.

\end{document}